\documentclass{andromedaone}       

\journal{BSM}
\vol{2021}
\jyear{Egypt ~~~~~~~ UdeM-GPP-TH-21-287}
\pages{Zewail City of Science, Technology and Innovation} 

\received{xx January 2018}
\published{xx March 2018}

\def\beq{\begin{equation}}
\def\eeq{\end{equation}}
\def\bea{\begin{eqnarray}}
\def\eea{\end{eqnarray}}

\newcommand{\nn}{\nonumber}

\def\nn{\nonumber}
\def\roughly#1{\mathrel{\raise.3ex\hbox
{$#1$\kern-.75em\lower1ex\hbox{$\sim$}}}}
\def\lsim{\roughly<}

\def\sla#1{\raise.15ex\hbox{$/$}\kern-.57em #1}

\def\ACP{A_{\rm CP}}
\def\cM{{\cal M}}

\begin{document}

\title{Rare Lepton-Number-Violating $W$ Decays at the LHC: CP
  Violation}

\author{David London,\auno{1}}
\address{$^1$Physique des Particules, Universit\'e de Montr\'eal, 1375
  Avenue Th\'er\`ese-Lavoie-Roux, Montr\'eal, QC, Canada H2V 0B3}

\begin{abstract}
Some models of leptogenesis involve a nearly-degenerate pair of heavy
Majorana neutrinos $N_{1,2}$ whose masses can be small, $O({\rm
  GeV})$. There can be heavy-light neutrino mixing parametrized by
$|B_{\ell N}|^2 = 10^{-5}$, which leads to the rare
lepton-number-violating decay $W^\pm \to \ell_1^\pm \ell_2^\pm
(q'{\bar q})^\mp$. With contributions to this decay from both $N_1$
and $N_2$, a CP-violating rate difference between the decay and its
CP-conjugate can be generated.  In this talk, I describe the prospects
for measuring such a CP asymmetry $\ACP$ at the LHC. I consider three
versions of the LHC -- HL-LHC, HE-LHC, FCC-hh -- and show that, for
$5~{\rm GeV} \le M_N \le 80~{\rm GeV}$, small values of the CP
asymmetry can be measured at $3\sigma$, in the range $1\% \lsim \ACP
\lsim 15\%$.
\end{abstract}

\maketitle

\begin{keyword}
  Lepton-number violation\sep $W$ decays at the LHC\sep CP violation\sep
  leptogenesis models\sep light sterile neutrinos
  \doi{10.1007/JHEP04(2021)021}
\end{keyword}

\bigskip
\noindent
Talk based on work done in collaboration with Fatemeh Najafi and Jacky Kumar, Ref.~\cite{Najafi:2020dkp}.

\section{Introduction}

One of the fundamental mysteries in particle physics -- indeed, in all
of physics -- is the origin of the baryon asymmetry of the universe
(BAU). The only thing we know for sure about the BAU is that its
generation requires the three Sakharov conditions: \\ (i)
baryon-number violation, (ii) CP violation, (iii) processes that take
place out of equilibrium \cite{Sakharov:1967dj}. One popular
explanation is leptogenesis. Here, a lepton-number asymmetry is
created through CP-violating, lepton-number-violating processes. This
is then converted to a baryon-number asymmetry via sphalerons
processes \cite{tHooft:1976rip, tHooft:1976snw}, which conserve $B-L$.

Another mystery is neutrino masses, which are known to be nonzero, but
very small.  What is the origin of these neutrino masses? And are
neutrinos Dirac or Majorana particles? If they are Majorana,
low-energy lepton-number-violating processes such as neutrinoless
double-beta decay may be observable.

A common scenario in leptogenesis models, which also touches the
question of neutrino masses, is the appearance of a pair of
nearly-degenerate heavy sterile neutrinos $N_1$ and $N_2$.
Leptogenesis can then be produceed through CP-violating decays of the
heavy neutrinos \cite{Pilaftsis:1997jf, Pilaftsis:2003gt}, or via
neutrino oscillations \cite{Akhmedov:1998qx, Canetti:2012kh}. We will
see both of these effects below.

In the seesaw mechanism \cite{GellMann:1980vs, Yanagida,
  Mohapatra:1979ia} with one left-handed (LH) and one right-handed
(RH, sterile) neutrino, the mass matrix takes the form
\beq
M = \left( \begin{array}{cc} 0&m_D\\m_D&m_R\\ \end{array} \right) ~,
\eeq
leading to

\beq
m_\nu = \frac{m_D^2}{m_R} ~~,~~~~ m_N = m_R ~.
\eeq
The standard choice for the entries in the mass matrix is $m_D \sim
m_t$, $m_R \sim 10^{15}$ GeV. But there are other possibilities, e.g.,
$m_D \sim m_e$, $m_R \sim 1$ TeV. 

With three LH and three RH neutrinos, there are more free parameters
in the mass matrix (three $m_D$s and three $m_R$s). A complete scan of
the parameter space reveals that it is possible to obtain three
ultralight neutrinos $\nu_i$ and three heavy Majorana neutrinos $N_i$,
with $N_1$ and $N_2$ nearly degenerate and with masses of $O({\rm
  GeV})$ \cite{Canetti:2014dka}.

The flavour and mass eigenstates are related via
\beq
\nu_\ell  = \sum_{j=1}^3 B_{\ell j} \nu_j + \sum_{i=1}^3 B_{\ell N_i} N_i ~.
\eeq
Here the $B_{\ell N_i}$ parametrize the heavy-light neutrino mixing.
The point is the following. With $B_{\ell N_i} \ne 0$, there are
$W$-$\ell$-$N_i$ couplings.  And if $M_N < M_W$, one can have the
decay $W^- \to \ell_1^- N_i$, with (i) $N_i \to \ell_2^- \ell_3^+
\nu_{\ell_3} ~,~ \ell_2^- (q' {\bar q})^+$ or (ii) $N_i \to \ell_2^+
\ell_3^- {\bar\nu}_{\ell_3}$.  Decays of type (i) are lepton-number
violating (LNV, $\Delta L = 2$), while decays of type (ii) are
lepton-number conserving (LNC, $\Delta L = 0$).  Searches for such
decays constrain the mixing parameters to be
\beq
|B_{\ell N}|^2 \le 10^{-5} ~~~ (\ell = e,\mu) ~,
\eeq
for 1 GeV $\le m_N \le$ 80 GeV \cite{Deppisch:2015qwa}.

The idea that there can be a pair of nearly-degenerate Majorana
neutrinos with masses of $O({\rm GeV})$ has led a number of authors to
examine the prospects for observing CP-violating LNV processes in the
decays of mesons \cite{Cvetic:2013eza, Cvetic:2014nla, Dib:2014pga,
  Cvetic:2015naa, Cvetic:2015ura, Cvetic:2020lyh, Godbole:2020doo,
  Zhang:2020hwj} and $\tau$ leptons \cite{Zamora-Saa:2016ito,
  Zamora-Saa:2019naq}. For example, the decay $B^\pm \to D^0
\ell_1^\pm \ell_2^\pm \pi^\mp$ is considered in
Ref.~\cite{Cvetic:2020lyh}. It occurs via $B^\pm \to D^0 W^{*\pm} (\to
\ell_1^\pm N_i)$, with $N_i \to \ell_2^\pm W^{*\mp} (\to \pi^\mp)$.

The key point here is that we can search for similar effects in the
decays of {\it real} $W$s at the LHC, in $W^- \to \ell_1^- \ell_2^-
(f' {\bar f})^+$. This decay has already been studied extensively as a
signal of LNV.  Here we push further and examine the prospects for
measuring CP violation in this decay.

As noted above, in $W^- \to \ell_1^- N_i$, if the $N_i$ decays
leptonically, the final state can be $\ell_1^- \ell_2^- \ell_3^+
\nu_{\ell_3}$ (LNV) or $\ell_1^- \ell_2^+ \ell_3^- {\bar\nu}_{\ell_3}$
(LNC). Since the final-state (anti)neutrino is not detected, these are
indistinguishable. However, we want to focus on pure LNV decays, so in
our study we consider only $W^- \to \ell_1^- \ell_2^- (q'{\bar
  q})^+$. A difference between the rates of $W^- \to \ell_1^- \ell_2^-
(q'{\bar q})^+$ and its CP-conjugate decay $W^+ \to \ell_1^+ \ell_2^+
(q'{\bar q})^-$ is a signal of CP violation.

\section{CP Violation -- Review}

\setcounter{equation}{4}

Suppose that the decay $W^- \to F$, where $F$ is the final state, has
two contributing amplitudes, $A$ and $B$:
\beq
A_{\rm tot} = A + B = |A| e^{i\phi_A} e^{i\delta_A} + |B| e^{i\phi_B} e^{i\delta_B} ~,
\eeq
where $\phi_{A,B}$ and $\delta_{A,B}$ are CP-odd and CP-even phases,
respectively. The CP asymmetry is
\bea
\ACP &=& \frac{BR(W^- \to F) - BR(W^+ \to {\bar F})}{BR(W^- \to F) + BR(W^+ \to {\bar F})} \nn\\
&=& \frac{2 |A| |B| \sin (\phi_A - \phi_B) \sin (\delta_A - \delta_B)}
{ |A|^2 + |B|^2 + 2 |A| |B| \cos (\phi_A - \phi_B) \cos (\delta_A - \delta_B)} ~. 
\label{ACP}
\eea
From this we see that a nonzero $\ACP$ requires the two contributing
amplitudes to have different CP-odd phases ($\phi_A - \phi_B \ne 0$)
{\it and} different CP-even phases ($\delta_A - \delta_B \ne 0$). In
addition, $\ACP$ is sizeable only when the two amplitudes are of
similar size ($|A| \sim |B|$).

In $W^- \to \ell_1^- \ell_2^- (q' {\bar q})^+$, the two amplitudes are
$W^- \to \ell_1^- {\bar N}_{1,2}$, with each of ${\bar N}_{1,2}$
decaying to $\ell_2^- (q' {\bar q})^+$. Here $\phi_1 = \arg[B_{\ell_1
    N_1} B_{\ell_2 N_1}]$ and $\phi_2 = \arg[B_{\ell_1 N_2} B_{\ell_2
    N_2}]$, so that $\phi_1 - \phi_2$ can be nonzero.

There are two sources of CP-even phases. First, the $N_i$
  propagator is proportional to
\bea
\frac{1}{(p_N^2 - M_{N_i}^2) + i M_{N_i} \Gamma_{N_i}}
&=& \frac{1}{\sqrt{(p_N^2 - M_{N_i}^2)^2 + M_{N_i}^2 \Gamma_{N_i}^2}} \, e^{i \eta_i} ~, \nn\\
{\rm with} ~~~~~
\tan\eta_i &=& \frac{- M_{N_i} \Gamma_{N_i}}{(p_N^2 - M_{N_i}^2)} ~.
\eea
As $N_1$ and $N_2$ do not have exactly the same mass, this leads to
$\eta_1 - \eta_2 \ne 0$. For example, if $\eta_1 = - \pi/2$ (i.e.,
$N_1$ is on-shell), then $|\eta_2| < \pi/2$. This is {\it resonant CP
  violation}.

Note also that, since the $N_i$ are nearly degenerate, the two
amplitudes are of similar size, so that $\ACP$ can be sizeable.

Second, there can be oscillations of heavy neutrinos. The time
evolution of a heavy $N_i$ mass eigenstate involves the factor $e^{-i
  E_i t}$, where $E_i$ is the energy of the $N_i$ in the rest frame of
the decaying $W$. Once again, since $M_{N_1} \ne M_{N_2}$, we have
$E_1 \ne E_2$, which gives different $e^{-i E_i t}$ factors. This is
another source of a CP-even phase difference, and can also lead to CP
violation.

\newpage

\section{$\cM(W^- \to \ell_1^- {\bar N}_i, N_i \to \ell_2^- W^{*+} (\to (q' {\bar q})^+)$}

\setcounter{equation}{7}

\begin{figure}[h]
\begin{center}
\includegraphics[width=0.4\textwidth]{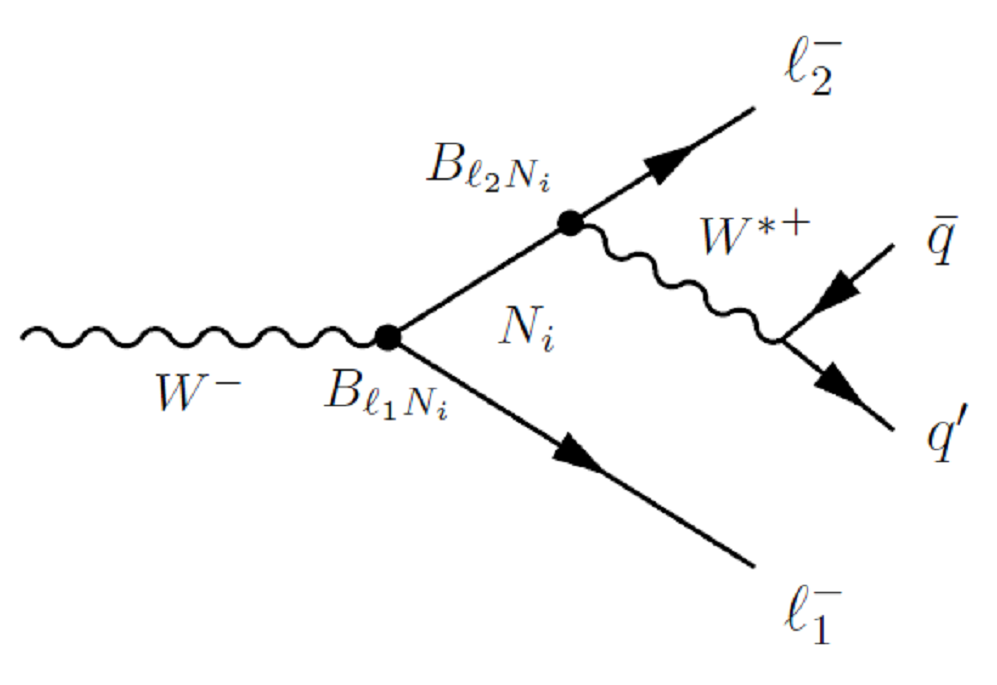}
\end{center}
\caption{Diagram for $W^- \to \ell_1^- \ell_2^- (q'{\bar q})^+$
  via an intermediate $N_i$. There is no arrow on the $N_i$ line
  because it is a Majorana particle and the decay is fermion-number
  violating.}
\label{Ndecayfig}
\end{figure}

The Feynman diagram for $W^- \to \ell_1^- \ell_2^- (q'{\bar q})^+$ via
an intermediate $N_i$ is shown in Fig.~\ref{Ndecayfig}. 

Because this decay receives contributions from $N_i = N_1$ and $N_2$,
and since the two neutrinos cannot be on shell simultaneously, we must
include the heavy neutrino propagator in the amplitude. In addition,
although the neutrino is produced as ${\bar N}_i$, it actually decays
as $N_i$, leading to the fermion-number-violating and LNV process $W^-
\to \ell_1^- \ell_2^- (q'{\bar q})^+$. This implies that (i) conjugate
fields will be involved in the amplitudes, and (ii) the amplitudes
will be proportional to the neutrino mass.

The full amplitudes are $\cM_i^{--} \equiv \cM(W^- \to \ell_1^- {\bar
  N}_i, {\bar N}_i \to N_i, N_i \to \ell_2^- W^{*+} (\to (q' {\bar
  q})^+)$.  Writing $\cM_i^{--} = \cM_i^{\mu\nu} \epsilon_\mu j_\nu$,
where $\epsilon_\mu$ is the polarization of the initial $W^-$ and
$j_\nu = \frac{g}{\sqrt 2} {\bar q} \gamma_\nu P_L q'$ is the current
of final-state particles to which $W^{*+}$ decays, we have
\bea
\cM_i^{\mu\nu} &=&
\bar \ell_1 \gamma^\mu P_L \left (\frac{g}{\sqrt 2} B_{\ell_1 N_i} \right ) N_i 
\times e^{-\Gamma_i t/2} e^{-iE_i t} \times
\bar \ell_2 \gamma^\nu P_L \left (\frac{g}{\sqrt 2} B_{\ell_2 N_i} \right ) N_i \nn\\
&\to&
\frac{ \frac{g^2}{2} \, B_{\ell_1 N_i} \, B_{\ell_2 N_i} \, M_i \, e^{-\Gamma_i t/2} \, e^{-iE_i t}}
     {p_N^2 - M^2_i + i \Gamma_i M_i} \,
L^{\mu\nu} ~,
\eea
where $L^{\mu\nu} = \bar \ell_1 \gamma^\mu \gamma^\nu P_R \ell_2^c$.
In the first line, the first term is the amplitude for $W^- \to
\ell_1^- {\bar N}_i$, the second term is the time dependence of the
$N_i$ state, and the third term is the amplitude for $N_i \to \ell_2^-
W^{*+}$. The $e^{-iE_i t}$ factor is due to the quantum-mechanical
evolution of the $N_i$ state (neutrino oscillations). The CP-odd phase
is found in $B_{\ell_1 N_i} B_{\ell_2 N_i}$, while the CP-even phase
arises from the $e^{-iE_i t}$ and $i \Gamma_i M_i$ factors.

The total amplitude is $\cM^{\mu\nu} = \cM_1^{\mu\nu} +
\cM_2^{\mu\nu}$. Writing $B_{\ell_1 N_1} B_{\ell_2 N_1} \equiv B_1
e^{i \phi_1}$ and $B_{\ell_1 N_2} B_{\ell_2 N_2} \equiv B_2 e^{i
  \phi_2}$, we have
\beq
\cM^{\mu\nu} = \frac{g^2}{2} \left(
\frac{M_1 \, B_1 \, e^{i \phi_1} \, e^{-\Gamma_1 t/2} \, e^{-iE_1 t}}{p_N^2-M_1^2+ i\Gamma_1 M_1}
+ \frac{M_2 \, B_2 \, e^{i \phi_2} \, e^{-\Gamma_2 t/2} \, e^{-iE_2 t}}{p_N^2-M_2^2+ i\Gamma_2 M_2} \right)
L^{\mu\nu} ~.
\eeq
Note that the two contributing amplitudes have different CP-odd phases
($\phi_1$ and $\phi_2$) and (two sources of) different CP-even phases
($i\Gamma_1 M_1$ vs.\ $i\Gamma_2 M_1$ and $e^{-iE_1 t}$ vs.\ $e^{-iE_2
  t}$). We therefore expect to find a CP asymmetry.

Using this expression, we (i) compute $|\cM^{\mu\nu}|^2$ using the
narrow-width approximation, (ii) integrate over time (our goal is not
the measurement of the neutrino oscillations), (iii) perform the
phase-space integrals, and (iv) construct $\ACP$.

\newpage

\section{$\ACP$}

\setcounter{equation}{9}

With the simplifying assumption that $B_1 = B_2$, we find
\beq
\ACP = \frac{ 2 (2 y - x) \sin \delta \phi}{ ( 1+x^2 ) ( 1 + 4y^2 ) + 2 ( 1-2xy ) \cos \delta \phi } ~,
\label{ourACP}
\eeq
where
\beq
x \equiv \frac{\Delta E}{\Gamma} ~~,~~~~ y \equiv \frac{\Delta M}{\Gamma} ~~,
~~~~{\hbox{with}}~~~~ x = y \, \frac{M_N}{M_W} ~.
\eeq
Comparing Eqs.~(\ref{ACP}) and (\ref{ourACP}), we see that $x$ and $y$
each play the role of the CP-even phase-difference term $\sin
(\delta_A - \delta_B)$. $x$ arises from neutrino oscillations (hence
the factor $\Delta E$), while $y$ is due to the neutrino propagator
($\Delta M$).

We note that $y$ is always present; $x$ is generally subdominant,
except for large values of $M_N$. Given that $|2y - x| \le |2y|$, this
implies that that, as $|x|$ increases, $\ACP$ decreases. We therefore
expect to see smaller CP-violating effects for larger values of $M_N$.

\begin{figure}[h]
\begin{center}
\includegraphics[width=0.4\textwidth]{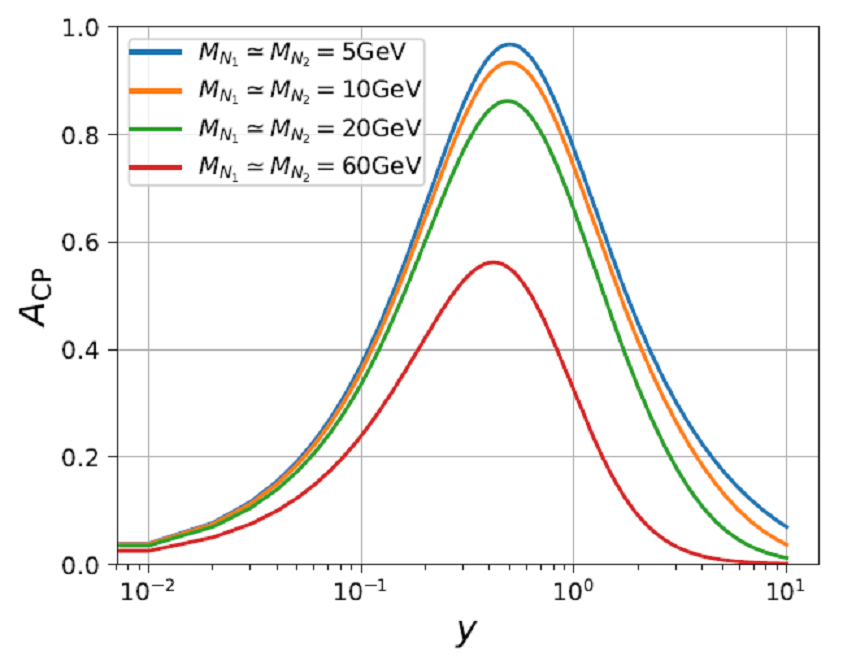}
\end{center}
\caption{Value of $\ACP$ as a function of $y$, for $\delta \phi
  = \pi/2$ and for various values of $M_N$. For negative values of
  $y$, $\ACP \to -\ACP$.}
\label{ACPfig}
\end{figure}

In order to estimate the potential size of $\ACP$, we set $\delta \phi
= \pi/2$.  In Fig.~\ref{ACPfig}, we plot $\ACP$ as a function of $y$,
for various values of $M_N$.

From this plot, we note the following features:
\begin{itemize}
  
\item Large values of $|\ACP|$ ($\ge 0.9$) can be produced for light
  $M_N$.

\item Maximal values of $|\ACP|$ are found when $y \simeq \pm\frac12$,
  with $|\ACP|$ decreasing for larger/smaller values of $|y|$.

\item As expected, the size of $|\ACP|$ decreases as $M_N$ increases,
  with $|\ACP|_{\rm max} < 0.6$ for larger values of $M_N$. (Even so,
  these values of $|\ACP|$ are not that small.)

\end{itemize}

Note in passing: the observation that CP violation is maximal when $y
\simeq \pm\frac12$ allows us to quantify how degenerate the
``nearly-degenerate heavy sterile neutrinos'' must be. Using $y \equiv
\Delta M/\Gamma$, we find that, for $M_N = 10$ GeV, $\Delta M =
O(10^{-14})$ GeV.

\section{Experimental Prospects}

\setcounter{equation}{11}

In order to measure $\ACP$, one has to compare $N_{--}$ (the number of
events of $W^- \to \ell_1^- \ell_2^- (q'{\bar q})^+$) and $N_{++}$
(the number of events of $W^+ \to \ell_1^+ \ell_2^+ (q{\bar q}')^-$).
However, one must also take into account the fact that, because $pp$
collisions are involved at the LHC, and because protons do not contain
an equal number of up- and down-type quarks and antiquarks, the number
of $W^-$ and $W^+$ bosons produced is not equal. This can be done by measuring
\beq
\ACP = = \frac{ R_W N_{--}^{pp} - N_{++}^{pp}}{ R_W N_{--}^{pp} + N_{++}^{pp}} ~,
\eeq
where $N_{--}^{pp}$ and $N_{++}^{pp}$ are the number of observed
events of $pp \to X W^- (\to \ell_1^- \ell_2^- (q'{\bar q})^+)$ and
$pp \to X W^+ (\to \ell_1^+ \ell_2^+ ({\bar q}' q)^-)$, respectively,
and 
\beq
R_W = \frac{\sigma(pp \to W^+ X)}{\sigma(pp \to W^- X)} ~,
\eeq
measured to be $R_W = 1.295 \pm 0.003~(stat) \pm 0.010~(syst)$ at
$\sqrt{s} = 13$ TeV \cite{Aad:2016naf}. Presumably, $R_W$ can be
measured with equally good precision (if not better) at higher
energies.

Now, given an $\ACP$, the number of events required to show it is nonzero
at $n\sigma$ is
\beq
N_{\rm events} = \frac{n^2}{A_{CP}^2 \, \epsilon} ~,
\eeq
where $\epsilon$ is the experimental efficiency. This can be turned
around: given $N_{\rm events}$, we can compute the smallest value of
$|\ACP|$ that can be measurable at $n\sigma$.

In our study, we consider three versions of the LHC: (i) the
high-luminosity LHC (HL-LHC, $\sqrt{s} = 14$ TeV), (ii) the
high-energy LHC (HE-LHC, $\sqrt{s} = 27$ TeV), and (iii) the future
circular collider (FCC-hh, $\sqrt{s} = 100$ TeV). We
implement the model in {\tt FeynRules} \cite{Degrande:2016aje,
  Alloul:2013bka} and use {\tt MadGraph} \cite{Alwall:2014hca} to
generate events. We take $|B_{\ell N}|^2 \le 10^{-5}$.

Note that $N_{\rm events}$ is not whole story. What we really want is
the number of {\it measurable} events. To be specific, we require that
the sterile neutrinos actually decay in the detector. With this in
mind, it is necessary to look at the $N$ lifetime and determine what
percentage of the heavy neutrinos actually decay in the detector. This
was done by the CMS Collaboration in its search for $W^- \to \ell_1^-
\ell_2^- (f'{\bar f})^+$ \cite{Sirunyan:2018mtv}. They found that, for
$M_N = 1$ GeV, 5 GeV and 10 GeV, the multiplicative reduction factor
was $10^{-3}$, 0.1 and $\simeq 1$, respectively.

In its searches for heavy Majorana neutrinos at the $\sqrt{s} = 8$ TeV
LHC using the final state $\ell_1^- \ell_2^- j j$
\cite{Khachatryan:2015gha, Khachatryan:2016olu}, the CMS Collaboration
found that their overall efficiency was $\sim 1$\%. Using this
efficiency in our estimates, we obtain the results given in Table
\ref{ACPminTable}.

\begin{table}[h!]
\tbl{Minimum value of $\ACP$ measurable at $3\sigma$ at the
  HL-LHC ($\sqrt{s} = 14$ TeV), HE-LHC ($\sqrt{s} = 27$ TeV) and
  FCC-hh ($\sqrt{s} = 100$ TeV). Results are given for $M_N = 5$ GeV
  (reduction factor $= 0.1$), $M_N = 10$ GeV (no reduction factor),
  and $M_N = 50$ GeV (no reduction factor).\label{ACPminTable}}{%
\begin{tabular}{ |c|c|c|c|  } \hline
  \multicolumn{4}{|c|}{Minimum $\ACP$ measurable at $3\sigma$} \\
\hline
Machine & $M_N = 5$ GeV & $M_N = 10$ GeV & $M_N = 50$ GeV \\
\hline
HL-LHC & 15.0\% & 4.8\% & 7.4\% \\
HE-LHC & 5.1\% & 1.6\% & 2.5\% \\
FCC-hh & 2.1\% & 0.7\% & 1.0\% \\
\hline
\end{tabular}}
\end{table}

We note that
\begin{itemize}

\item As LHC increases in energy and integrated luminosity, smaller
  values of $\ACP$ are measurable.

\item At a given machine, the measurable $\ACP$ decreases as $M_N$
  increases. (But there is a reduction factor due to the $N$ lifetime
  for small $M_N$.)

\item The most promising results are for $M_N = 10$ GeV, but in all
  cases reasonably small values of $\ACP$ can be probed.

\end{itemize}

\section{Summary}

In many leptogenesis models, a lepton-number asymmetry arises through
CP-violating decays of a pair of nearly-degenerate heavy neutrinos
$N_1$ and $N_2$. What is particularly intriguing is that the masses of
$N_{1,2}$ can be small, $O({\rm GeV})$.

In general, there can be a (small) heavy-light neutrino mixing. This
leads to LNV processes at the LHC such as $W^\pm \to \ell_1^\pm
\ell_2^\pm (q'{\bar q})^\mp$. A CP-violating rate asymmetry $\ACP$
between the $W^-$ and $W^+$ decays can arise due to the interference
of the $N_1$ and $N_2$ contributions. The different $W$-$\ell$-$N_1$
and $W$-$\ell$-$N_2$ couplings produce the CP-odd phase difference;
The CP-even phase difference is generated via propagator effects or
oscillations of the heavy neutrinos.

If such an LNV decay were observed, this would of course be very
exciting. But the next step would be to try to understand the
underlying origin of the decay. One important piece of information
would be to look at CP violation in the decay, and this is what we
have studied.

We consider $5~{\rm GeV} \le M_N \le 80~{\rm GeV}$ and examine three
versions of the LHC: (i) HL-LHC ($\sqrt{s} = 14$ TeV), (ii) HE-LHC
($\sqrt{s} = 27$ TeV), (iii) FCC-hh ($\sqrt{s} = 100$ TeV).  The most
promising result is for the FCC-hh with $M_N = 10$ GeV. Here $\ACP =
O(1\%)$ is measurable. But even in the worst case, the HL-LHC with
$M_N = 5$ GeV, an $\ACP = O(10\%)$ can be measured.

\section*{Acknowledgements}

This work was financially supported by NSERC of Canada.

\newpage

\bibliographystyle{unsrt}

\end{document}